\documentclass[12pt]{iopart}
\usepackage{iopams}
\usepackage{graphicx}
\usepackage[dvips]{color}
\usepackage{amssymb, amsfonts, amsbsy}

\begin{document}
%------------------------------------------------
\title[Cosmodynamics: Energy conditions, Hubble bounds, density bounds, ...]    
{Cosmodynamics:  Energy conditions, Hubble bounds, density bounds, time and distance bounds}
%------------------------------------------------
\author{C\'eline Catto\"en and Matt Visser}
%------------------------------------------------
\address{School of Mathematics, Statistics, and Computer Science, 
Victoria University of Wellington, PO Box 600, Wellington, New Zealand}
%------------------------------------------------
\ead{celine.cattoen@mcs.vuw.ac.nz, matt.visser@mcs.vuw.ac.nz}
%------------------------------------------------
\begin{abstract}
%------------------------------------------------

\vskip 0.50cm

We refine and extend a programme initiated by one of the current authors [Science  {\bf 276} (1997) 88; Phys.~Rev.~{\bf D56} (1997) 7578] advocating the use of the classical energy conditions of general relativity  in a cosmological setting to place very general bounds on various cosmological parameters. We show how the energy conditions can be used to bound the Hubble parameter $H(z)$, Omega parameter $\Omega(z)$, density $\rho(z)$, distance $d(z)$, and lookback time $T(z)$ as (relatively) simple functions of the redshift $z$, present-epoch Hubble parameter $H_0$, and present-epoch Omega parameter $\Omega_0$. We compare these results with related observations in the literature, and confront the bounds with the recent supernova data.

\vskip 0.50cm
\noindent
  Dated: 10 December 2007; \LaTeX-ed \today
%------------------------------------------------
\end{abstract}
%------------------------------------------------
%\pacs{arXiv:1207.nnnn}
%------------------------------------------------
\maketitle

%---------------------------------------
% definitions
%------------------------------------------------
\newtheorem{theorem}{Theorem}
\newtheorem{corollary}{Corollary}
\newtheorem{lemma}{Lemma}
%------------------------------------------------
\def\d{{\mathrm{d}}}
\def\implies{\Rightarrow}

\def\NEC{{_\mathrm{NEC}}}
\def\WEC{{_\mathrm{WEC}}}
\def\SEC{{_\mathrm{SEC}}}
\def\DEC{{_\mathrm{DEC}}}

%------------------------------------------------

%-----------------------------------------------------------------------------
%\def\lint{\hbox{\Large $\displaystyle\int$}} %needs \usepackage{amssymb}
%\def\hint{\hbox{\Huge $\displaystyle\int$}}  %needs \usepackage{amssymb}
%-----------------------------------------------------------------------------
\def\eg{{\it e.g.}}
\def\ie{{\it i.e.}}
\def\etc{{\it etc.}}
\def\sign{{\hbox{sign}}}
%-------------------------------------------------------------------------
\def\eof{\Box}
%-------------------------------------------------------------------------
\newenvironment{warning}{{\noindent\bf Warning: }}{\hfill $\eof$\break}
%-------------------------------------------------------------------------

%------------------------------------------------------------------------------------------
\section{Introduction}
%------------------------------------------------------------------------------------------
Some 10 years ago one of the current authors initiated a programme of using the classical energy conditions of general relativity to place very general bounds on various cosmological parameters~\cite{science, flow, mg8}. In that early work, attention was mainly focussed on the energy density $\rho(z)$ and lookback time $T(z)$. Since then, the classical energy conditions have (on the one hand) seen continued use in studying issues such as the minimal requirements for cosmological bounces~\cite{bounce, tolman} and other ``cosmological milestones''~\cite{milestones}, and (on the other hand) have seen further applications to bounding cosmological distances $d(z)$~\cite{santos1,santos2}, and lookback time $T(z)$~\cite{santos3}. In the current article we shall try to draw these various threads together and establish several simple and rugged energy-condition-induced bounds on cosmological parameters. For some generic cosmological parameter, say represented by $X(z)$, we shall seek bounds of the form
\begin{equation}
X(z) \gtrless X_\mathrm{bound} \equiv X_0  \; f(\Omega_0,z),
\end{equation}
where $X_0$ is the value of $X(z)$ at the present epoch, the direction of the inequality may depend both on the bound being considered and the redshift region of interest, and $f(\Omega_0,z)$ is some dimensionless function to be determined.

There is (at least) one important caveat: It should be kept clearly in mind that the classical energy conditions are \emph{not} fundamental physics --- in fact the classical energy conditions are known to be violated by quantum effects~\cite{scale, gvp, cosmo99, implications}, at least to some extent, and so the energy conditions should always be viewed provisionally --- as a way of characterizing whether or not a certain situation is describable by ``normal'' physics~\cite{cosmo99, implications}.

%------------------------------------------------------------------------------------------
\section{Basic formulae}
%------------------------------------------------------------------------------------------
In standard cosmology, one assumes the cosmological principle, that is, our universe is isotropic and homogeneous on large scales. This assumption leads one to consider cosmological spacetimes of the idealized FLRW form~\cite{MTW, Weinberg, Wald, Peebles, Carroll, Hartle}:
\begin{eqnarray}
\d s^{2}&=&-\d t^{2}+a(t)^{2}\left\{\frac{\d r^{2}}{1-kr^{2}}+r^{2}\;[\d\theta^{2}+\sin^2\theta\; \d\phi^{2}]\right\}.
\end{eqnarray}
If we further assume that gravitational interactions at large scales are described by general relativity, we can use the Friedmann equations that relate the total density $\rho$ and the total pressure $p$ to a function of the scale factor $a$ and its time derivatives. Indeed, in units where $8\pi G_N=1$, but explicitly retaining the speed of light $c$, we have\footnote{Note that we are now specifically assuming Friedmann dynamics for the universe, one is thus explicitly stepping outside the ``cosmographic'' or ``cosmokinetic'' framework of~\cite{Chiba,  Sahni, Jerk,  Jerk2,  Blandford, Blandford0, Cattoen1, Cattoen2, Seikel}.    } 
\begin{eqnarray}
 \rho(t)&=&3\left(\frac{\dot a^{2}}{a^{2}}+\frac{k \,c^2}{a^{2}}\right),
\label{Friedmann1}
\end{eqnarray}
\begin{eqnarray}
p(t)&=&-2\,{\ddot a\over a}-{\dot a^2\over a^2} - {k \,c^2 \over a^2},
\label{Friedmann2}
\end{eqnarray}
\begin{eqnarray}
\rho(t)+3\,p(t)&=&-6\,{\ddot a\over a}.
\label{Friedmann3}
\end{eqnarray}
The classical energy conditions of general relativity, to the extent that one believes that they are a useful guide~\cite{cosmo99,implications}, allow one to deduce physical constraints on the behaviour of matter fields in strong gravitational fields or cosmological geometries. These conditions can most easily be stated in terms of  the components of the stress energy tensor $T^{\hat\mu \hat\nu}$ in an orthonormal frame. Ultimately, however, constraints on the stress-energy are converted, via the Einstein equations, to constraints on the spacetime geometry --- in particular in a FLRW spacetime one is ultimately imposing conditions on the scale factor and its time derivatives. For a perfect fluid cosmology, and in terms of pressure and density, the so-called  \emph{Null}, \emph{Weak}, \emph{Strong} and \emph{Dominant} energy conditions reduce to~\cite{Lorentzian}:
\begin{description}

\item[NEC:] $\rho+p \geq 0$. \\
In view of the Friedmann equations this then reduces to
\begin{equation} 
\label{nec}
-\frac{\ddot{a}}{a}+\frac{\dot{a}^2}{a^2}+\frac{k\, c^2}{a^2} \geq 0;
\qquad\hbox{that is}\qquad
\frac{\ddot{a}}{a} \leq \frac{\dot{a}^2}{a^2}+\frac{k\, c^2}{a^2}.
\end{equation}

\item[WEC:] This specializes to the NEC plus $\rho\geq0$.\\
This then reduces to the NEC plus the condition
\begin{equation} 
\label{wec}
\dot a^2 + k\, c^2 \geq 0.
\end{equation}
This condition is \emph{vacuous} for $k\in\{0,+1\}$ and only for $k=-1$ does it convey even a little information.

\item[SEC:] This specializes to the NEC plus $\rho+3p\geq0$.\\
This then reduces to the NEC plus the deceleration condition
\begin{equation}
\frac{\ddot a}{a} \leq 0. \label{sec}
\end{equation}

\item[DEC:] $\rho\pm p \geq 0$.\\
This reduces to the NEC plus the condition
\begin{equation} 
\label{dec}
\frac{\ddot a}{a} \geq -2\left(\frac{\dot a^2}{a^2}+\frac{k \, c^2}{a^2}\right).
\end{equation}
\end{description}
Note particularly that the condition (\ref{sec}) is independent of the space curvature $k$.
Now, DEC implies WEC implies  NEC, and SEC implies NEC, but otherwise the NEC,  WEC, SEC, and DEC are mathematically independent assumptions. In particular, the SEC does \emph{not} imply the WEC. Violating the NEC implies violating the DEC, SEC, and WEC as well~\cite{Lorentzian}.

Note that ideal relativistic fluids satisfy the DEC, and certainly all the known forms of normal matter encountered in our solar system satisfy the DEC.  With sufficiently strong self-intereactions relativistic fluids can be made to violate the SEC (and DEC); but classical relativistic fluids always seem to satisfy the NEC. Most classical fields (apart from non-minimally coupled scalars) satisfy the NEC. Violating the NEC seems to require either quantum physics (which is unlikely to be a major contributor to the overall cosmological evolution of the universe) or non-minimally coupled scalars (implying that one is effectively adopting some form of scalar-tensor gravity).

Using this dynamical formulation of the energy conditions, Santos \emph{et al.}~\cite{santos1} derive some bounds, for the special case $k=0$, on the luminosity distance $d_L$ of supernovae, and then contrast this with the legacy~\cite{legacy,legacy-url} and gold~\cite{gold06} datasets. In reference~\cite{santos2} bounds on the distance modulus are presented for general values of $k\in\{-1,0,+1\}$, while in reference~\cite{santos3} they concentrate on the lookback time. Herein, we shall use a similar but distinct approach to obtain rugged bounds on the Hubble parameter, the Omega parameter, the density, the lookback time, and on the various distance scales defined previously in~\cite{Cattoen1, Cattoen2} \emph{for all values of k space curvature} $\in\{-1,0,+1\}$.

%------------------------------------------------------------------------------------------
\section{Bounds on the Hubble parameter}
%------------------------------------------------------------------------------------------
The energy conditions translate, in a FLRW setting, into the inequalities (\ref{nec}), (\ref{wec}), (\ref{sec}), and (\ref{dec}), from which we deduce bounds on the Hubble function $H(z)$ in terms of the Hubble parameter $H_0$, the Omega parameter $\Omega_0$, and the $z$-redshift. 
\begin{description}
%------------------
\item[NEC:] 
Using inequality (\ref{nec}) we obtain:
\begin{eqnarray}
\frac{\dot{a}}{a}\; \frac{\d}{\d a}\left( \frac{\dot{a}}{a} \right) & \leq & \frac{k\,c^2}{a^3},
\end{eqnarray}
which can be integrated to yield
\begin{eqnarray}
\int_a^{a_0}  \frac{\d}{\d a}\left( \frac{1}{2}\; \left( \frac{\dot{a}}{a} \right)^2 \right) \; \d a 
& \leq &
\int_a^{a_0} \frac{k\; c^2}{a^3} \; \d a.
\end{eqnarray}
That is
\begin{equation}
H_0^2 - H(z)^2 \leq -{k \,c^2}\; \{ a_0^{-2} - a^{-2} \}
\end{equation}
Now using
\begin{eqnarray} 
\label{a0-z_relation}
\frac{a_0}{a}&=&1+z,
\end{eqnarray}
and the relation
 \begin{eqnarray} 
 \label{Omega0_definition}
\Omega_0 &=&1+\frac{kc^2}{a_0^2\, H_0^2},
\end{eqnarray}
after a few rearrangements  we obtain the bound:
\begin{eqnarray}  
H(z) & \geq &H_\NEC\;\equiv\;H_0 \sqrt{ \Omega_0  + \left[1- \Omega_0 \right] (1+z)^2 }.  
\label{H_NEC}
\end{eqnarray}
In order to obtain this inequality we have assumed that $z>0$, so that one is looking into the past. When looking into the future, $z<0$, the inequality is reversed.\footnote{Note that looking back into the past $z>0$, with $z=\infty$ corresponding to the big bang. In contrast, looking forward into the future $z<0$, with $z=-1$ corresponding to infinite expansion~\cite{Cattoen1,Cattoen2}.}

\paragraph{Technical point:} Existence of $\sqrt{ \Omega_0  + \left[1- \Omega_0 \right] (1+z)^2 }$ in the $\mathbb{R}$ domain.

The expression
\begin{eqnarray}
\Omega_0  + \left[1- \Omega_0 \right] (1+z)^2  &> & 0,
\end{eqnarray}
holds for all values of $z$, if and only if, $ \Omega_0 \leqslant 1$.
In contrast, note that when $\Omega_0 > 1$, there exists a $z$ value for which the expression in the square root becomes negative or zero. This specific value of the $z$-redshift is given by:
\begin{eqnarray}
z_{\NEC} & =& \sqrt{\frac{\Omega_0}{\Omega_0 -1}}-1.
\end{eqnarray}
Note that at $z=z_{\NEC}$, we get $H_{\NEC}(z_{\NEC})=0$. Also note that $z_\NEC$ is positive as long as $\Omega_0$ is positive. Nothing unusual need happen to the universe itself at $z_{\NEC}$, it is only the \emph{bound} that loses its predictive usefulness. In practice, given that current observational estimates are
\begin{eqnarray}
\Omega_0 &=&   1.02\pm0.02      \qquad \hbox{(PDG 2004~\cite{pdg04})},
\\
\Omega_0 &=&   1.003{\textstyle{+0.013\atop -0.017}}      \qquad \hbox{(PDG 2006~\cite{pdg06})},
\end{eqnarray}
let us point out that $z_{\NEC}(\Omega_0=1.04)=4.1$ and $z_{\NEC}(\Omega_0=1.01)=9.0$. So given current observational estimates of $\Omega_0$, the fact that the highest-$z$ supernovae seen to date have $z\lesssim 2$, and the fact that we are expected to run out of galaxies by the time we reach $z\lesssim 7$, the limitations associated with $z_{\NEC}$ are unlikely to be significant in any realistic setting.

Overall, the bound on the Hubble function (\ref{H_NEC}) is valid for $\Omega_0 \leq 1, \; \forall \; z \in \; [0,+\infty)$, and for $\Omega_0 > 1$ under the condition that $z\; \in [0,z_{\NEC} ]$.

%---------------
\item[WEC:]  From inequality (\ref{wec}), we can deduce that 
\begin{equation}  \label{H_WEC1}
\hbox{for } k=-1, \qquad \dot{a} \; \leq \; \sqrt{-k}\;c.
\end{equation}
 To obtain a relation on the Hubble function, we divide equation (\ref{H_WEC1}) by $a$, and obtain the rather weak bound:
 \begin{eqnarray}  \label{H_WEC}
H(z) & \geq & H_{\WEC} \equiv H_0 \; ( 1+z)\; \sqrt{1-\Omega_0} \qquad \forall \; \Omega_0\in(0,1).
\end{eqnarray}
We have assumed that $z>0$ in inequality (\ref{H_WEC}) so that one is looking into the past. When looking into the future $z<0$, the inequality is reversed. Note that this bound is only valid $\forall \; \Omega_0 \in(0,1)$ and $\forall \; z>0$.
\paragraph{Important remark:} Note that as long as $\Omega_0>0$ we have:
\begin{equation}
H_{\NEC} \geq H_{\WEC}. 
\label{compare_NEC_WEC}
\end{equation}
That is, the WEC really does not give us anything extra beyond the statement that $\Omega_0$ is positive.

%---------------
\item[SEC:]  From inequality (\ref{sec}), we deduce that 
\begin{equation}  \label{H_SEC1}
\forall \; a\; <\; a_0 \qquad \frac{1}{\dot{a}} \; \leq \;\frac{1}{H_0\,a_0}.
\end{equation}
 Further, to obtain a relation on the Hubble function and Hubble parameter, we multiply equation (\ref{H_SEC1}) by $a$, and we obtain the bound:
 \begin{eqnarray}  \label{H_SEC}
H(z) & \geq & H_{\SEC}\equiv H_0 \; ( 1+z).
\end{eqnarray}
We have assumed that $z>0$ in inequality (\ref{H_SEC}) so that one is looking into the past. When looking into the future $z<0$, the inequality is reversed. Note that this bound can also be found in~\cite{santos1, santos2}. 

%---------------
\item[DEC:] To satisfy this energy condition, the NEC must hold as well as inequality (\ref{dec}). We use the same approach as for the NEC, rewriting (\ref{dec}) as:
\begin{eqnarray}
\frac{\d (a^2 \dot{a})}{\d t} +2kc^2a & \geq & 0,
\end{eqnarray}
that is,
\begin{eqnarray}
\frac{\d a}{ \d t}\frac{\d}{\d a}(a^2\dot{a})+2kc^2a & \geq & 0.
\end{eqnarray}
Multiplying by $a^2$, this inequality leads to
\begin{eqnarray}
\frac{\d}{\d a} \left[ \frac{1}{2}\left( a^2 \,\dot{a}\right)^2+\frac{k\,c^2}{2}\;a^4 \right] & \geq & 0 \qquad \forall \; a. 
\end{eqnarray}
Integrating, we can deduce the new inequality,
\begin{eqnarray}
\forall \; a < a_0 \qquad  \left( a^2 \, \dot{a}\right)^2+k\,c^2\,a^4 & \leq  \left( a_0^2 \dot{a_0}\right)^2+k\,c^2\,a_0^4 .
\end{eqnarray}
Now, we multiply or divide appropriately by some combination of $a$ and $a_0$ to force the appearance of the Hubble function $H(z)$ and the Hubble parameter $H_0$. We also use equations (\ref{a0-z_relation}) and (\ref{Omega0_definition}) and substitute, leading to:   
\begin{eqnarray}
H(z) &\leq &    H_{\DEC} \;\equiv\;  H_0 \; (1+z)  \sqrt{1+ \Omega_0\; \left[ (1+z)^{4} - 1\right] } , \label{H_DEC}
\end{eqnarray}
Again, we have assumed that $z>0$ in inequality (\ref{H_DEC}) so that one is looking into the past. When looking into the future $z<0$, the inequality is reversed.

\emph{Thus the DEC is satisfied if and only if:
\begin{equation}
H_{\NEC} \; \leq \; H(z) \; \leq \; H_{\DEC},
\end{equation}
where $H_{\NEC}$ and $H_{\DEC}$ are defined respectively in equations (\ref{H_NEC}), and (\ref{H_DEC}).}

\paragraph{Technical point:} Existence of $\sqrt{1+ \Omega_0\; \left[ (1+z)^{4} - 1\right] } $  in the $\mathbb{R}$ domain: 

The expression
\begin{eqnarray}
1+ \Omega_0\; \left[ (1+z)^{4} - 1\right]   & > & 0,
\end{eqnarray}
holds, for all values of $z>-1$ (and hence $z \geq 0$) if $\Omega_0 \; \in \; (0,1)$. However, note that when $\Omega_0 \geqslant 1$, there exists a $z$ value for which the expression in the square root becomes negative or null. This specific value of the $z$-redshift is given by:
\begin{eqnarray}
z_{\DEC} & =&  \left(\frac{\Omega_0-1}{\Omega_0 }\right)^{1/4}-1.
\end{eqnarray}
Note that $z_{\DEC}$ is always negative so it is never a problem when looking back into the past. In fact $z_{\DEC}(\Omega_0=1.04)=-0.56$ and $z_{\DEC}(\Omega_0=1.01)=-0.68$ are well into the future.

Thus the bound on the Hubble function (\ref{H_DEC}) is valid for $\Omega_0 \in (0,1), \; \forall \; z \geq -1$, and for $\Omega_0 \geqslant 1$ under the condition that $z\; \in [z_{\DEC},+\infty ]$. If we are only interested in looking into the past, then the DEC bound holds for $\Omega_0>0$ and $z>0$.
\end{description}

%------------------------------------------------------------------------------------------
\section{Bounds on the distance scales}
%------------------------------------------------------------------------------------------

In order to obtain bounds on the various distance scales, it is enough to obtain a bound on Peebles' angular diameter distance~\cite{Peebles} and then use the different relations between the various distance scales presented in~\cite{Cattoen1, Cattoen2}.\footnote{Peebles' angular diameter distance is equal to Weinberg's proper motion distance~\cite{Weinberg}, and is also equal to D'Inverno's version of luminosity distance~\cite{dInverno}. Details on how the various distance scales are inter-related can be found in~\cite{Cattoen1, Cattoen2}.}   We choose to work primarily with Peebles' angular diameter distance because it minimizes the number of factors of $(1+z)$ occurring in the various formulae.  Peebles' angular diameter distance is generally defined in its exact form as~\cite{Peebles}:
\begin{eqnarray}
d_P(z) &=& a_0 \; 
\sin_k \left\{ {c\over H_0 \, a_0} \int_0^z {H_0\over H(z)} \; \d z \right\},
\end{eqnarray} 
where
\begin{eqnarray} \label{sink}
\sin_k(x)& =& 
\left\{ 
\begin{array}{ll}
       \sin(x), & k=+1;\\
        x,       & k=0;\\
        \sinh(x), & k=-1.\\
\end{array} \right.
\end{eqnarray} 
By changing variables and adopting definitions as in equations (\ref{a0-z_relation}) and (\ref{Omega0_definition}), we can rewrite Peebles' angular diameter distance in an alternative exact general form, $\forall \;z \in [-1,+\infty)$ and $\forall$ fixed $\Omega_0$:\footnote{Another notation that is sometimes used is $\Omega_k = 1-\Omega_0$, so that $k = -\hbox{sign}(\Omega_k)$.}
\begin{eqnarray}   \label{dPz_exact}
d_P(z) &=& \frac{c}{H_0 } \; 
\frac{\sinh \left[ \sqrt{1-\Omega_0} \int_0^z {H_0\over  H(z)} \; \d z \right] }{ \sqrt{1-\Omega_0}}  ,
\end{eqnarray} 
where we note
\begin{equation}
\Omega_0 \qquad \left\{
\begin{array}{ll}
      > 1, & \quad k=+1;\\
        =1, & \quad k=0;\\
     <1,  & \quad k=-1.\\
\end{array} \right.
\end{equation}
Observe  that by continuity of the functions $\sin x /x $ and $\sinh x/x$ as $x \to 0$, the function $d_P(z)$ is also continuous as $\Omega_0 \to 1^{\pm}$.
For convenience, from equation (\ref{dPz_exact}), the angular diameter distance is given by
\begin{eqnarray}  
d_P(z) &=& \frac{c}{H_0 } \; \frac{\sinh \left[ \sqrt{1-\Omega_0} \; J  \right] }{ \sqrt{1-\Omega_0}},
\end{eqnarray} 
where $J$ is the integral defined by
\begin{eqnarray}  \label{J}
J&=& \int_0^z {H_0\over  H(z)} \; \d z \;= \; H_0 \, a_0 \,\int_a^{a_0} \frac{\d a}{a\,\dot{a}}.
\end{eqnarray}
The procedure now is as follows: The energy conditions provide bounds on $H(z)$, which allow us to obtain a bound on the integral $J$. Then provided the function $\sin_k$ is monotonic on the interval $z \in [0,+ \infty]$, (or at least some sub-interval $z \in [0, z_\mathrm{max}]$\,), we can derive a bound on the angular diameter distance on this same domain.

\begin{description}
%----------------------
\item[NEC:] The null energy condition gives a bound on $H$ in equation (\ref{H_NEC}) leading to the inequality,
\begin{eqnarray}
J &=& \int_0^z {{H_0\over H(z)} \; \d z } \; \leq \; J_{\NEC}\;\equiv\; \int_0^z \frac{\d z}{ \sqrt{ \Omega_0  + \left[1- \Omega_0 \right] (1+z)^2 }} .
\end{eqnarray}
We integrate, and substitute the resulting bound back into the angular diameter distance. In the general case, we obtain the bound 
 \begin{eqnarray}  
& d_P(z)\;  \leq \; d_{P_{\NEC}}=\frac{c}{H_0\; \Omega_0} \; 
\left[ 1+z-\sqrt{\Omega_0+(1-\Omega_0)\,(1+z)^2} \right];
\label{dP_NEC}
\\
&
\qquad\qquad \left\{
\begin{array}{ll}
     \forall \; \Omega_0 \; \leq \; 1, \; \forall \; z \; \in [0,+\infty] ;\\
     \forall \;  \Omega_0 \; > \;  1, \; \forall \; z \; \in [0,z_{\NEC}).\\
\end{array} \right.  &  
\nonumber
\end{eqnarray} 
Note that as $\Omega_0 \to 1$ ($k=0$), we have
\begin{eqnarray}
d_{P}(z) & \leq & d_{L_{\NEC}}= \frac{c\; z}{H_0 } \qquad \Omega_0 = 1; \qquad \forall \; z \in [0,+\infty].
\end{eqnarray}
so we find the same particular result as in~\cite{santos1}, that is,
\begin{eqnarray}
d_{L}(z) & \leq & d_{L_{\NEC}}=\frac{c\; z\; (1+z)}{H_0 } \qquad \Omega_0=1; \qquad \forall \; z \in [0,+\infty].
\end{eqnarray}
Equation (\ref{dP_NEC}) is the more general case, valid for all values of $k\in\{-1,0,+1\}$. (Equation (15) of reference~\cite{santos2} can be viewed as an intermediate step in deriving equation  (\ref{dP_NEC}) above.)

Note that as $\Omega_0 \to 1$, equation (\ref{dP_NEC}) can be developed in a Taylor series as
\begin{eqnarray}
d_{P_{\NEC}}(z) & =&  \frac{c\; z}{ H_0} +\frac{c\; z^2}{2 \; H_0} \left( \Omega_0-1\right) + O\left([ \Omega_0-1]^2\right).
\end{eqnarray}
If instead one performs a low-redshift expansion, then for general $\Omega_0$
\begin{equation}
d_{P_{\NEC}}(z)  = {c z\over H_0} \left\{ 1 + {(\Omega_0-1)\; z\over2} + \mathcal{O}(z^2) \right\}.
\end{equation}

%----------------------
\item[WEC:] The weak energy condition gives a (weak) bound on $H(z)$ in equation (\ref{H_WEC}) only for $\Omega_0 \in(0,1)$ leading to the inequality,
\begin{eqnarray}
J &=& \int_0^z {{H_0\over H(z)} \d z } \; \leq \; J_{\WEC}\;\equiv\; \int_0^z \frac{\d z}{ \sqrt{ 1-\Omega_0} \, (1+z)} .
\end{eqnarray}
We integrate, and substitute the resulting bound back into the angular diameter distance. We obtain
 \begin{eqnarray}  
& d_P(z)\;  \leq \; d_{P_{\WEC}}=\frac{c}{H_0\; \sqrt{1-\Omega_0}} \; \frac{z \left( 2+z\right)}{\left( 1+z\right)};
\label{dP_WEC}
\\
&
\qquad    \forall \; \Omega_0 \; \in \; (0,1), \; \forall \; z \; \in [0,+\infty].
\nonumber
\end{eqnarray} 
\paragraph{Important remark:} Note that 
\begin{equation}
d_{P_{\NEC}} \leq d_{P_{\WEC}}.
\end{equation}
Thus the bound $d_{P_{\WEC}}$ is not very useful.

%--------------
\item[SEC:] This energy condition gives a bound on $H(z)$ in (\ref{H_SEC}), and therefore 
\begin{eqnarray}
J & \leq &J_{\SEC}\equiv \int_0^z \frac{\d z}{ 1+z } = \ln(1+z).
\end{eqnarray}
In the general case, we obtain the bound on the angular diameter distance (\emph{cf}. equation (17) of~\cite{santos2}):
 \begin{eqnarray}  
& d_P(z)\;  \leq \;d_{P_{\SEC}}(z) = \frac{c}{H_0} \; 
\frac{  \sinh \left[ \sqrt{1-\Omega_0}  \; \ln \left( 1+z \right)  \right] }{ \sqrt{1 - \Omega_0 }}; &  
 \label{dPSEC}
 \\
&
\qquad\qquad \left\{
\begin{array}{ll}
     \forall \; \Omega_0 \; \leq \; 1, \; \forall \; z \; \in [0,+\infty] ;\\
     \forall \;  \Omega_0 \; > \;  1, \; \forall \; z \; \in [0,z_{\mathrm{max}}].\\
\end{array} \right.  &  
\nonumber 
\end{eqnarray} 
In particular,
\begin{itemize}
\item For $k=-1$, that is $ \forall \; \Omega_0 <1$, and $ \forall \; z \in[0, + \infty]$,
\begin{equation}
d_{P_{\SEC}}(z) =  \frac{c}{ H_0} \; \frac{\left(1+z \right)^{\sqrt{1-\Omega_0}}-\left(1+z \right)^{ -\sqrt{1-\Omega_0} }} {2 \sqrt{1-\Omega_0}};
\end{equation}
\item  For $k=0$, that is when $ \Omega_0 =1$, and $ \forall \; z \in[0, +\infty]$ (\emph{cf}. the equivalent result in~\cite{santos1}),
\begin{equation}
d_{P_{\SEC}}(z) = \frac{c}{H_0} \; \ln \left( 1+z \right) ; 
\end{equation}
\item For $k=+1$, that is $ \forall \; \Omega_0 >1$, and $ \forall \; z \in[0, z_{\mathrm{max}}]$,
\begin{equation}
d_{P_{\SEC}}(z) = \frac{c}{H_0} \; 
\frac{  \sin \left[ \sqrt{\Omega_0-1} \; \ln \left( 1+z \right)   \right] }{ \sqrt{ \Omega_0-1 }},
\end{equation}
where\footnote{
If $J\leq J_\SEC$ then for the sine function we have $\sin(\sqrt{\Omega_0-1} \; J) \leq  \sin(\sqrt{\Omega_0-1} \; J_{\SEC})$, \emph{provided} that $0  \leq  \sqrt{\Omega_0-1}\; J_{\SEC} \leq \pi/2$, leading to the condition that $z \; \leq \; z_{\mathrm{max}}=\exp \left( \frac{\pi}{2\sqrt{\Omega_0-1}} \right) -1$.}
\begin{equation}
 z_{\mathrm{max}}=\exp \left( \frac{\pi}{2\sqrt{\Omega_0-1}} \right) -1.
\end{equation}
 In fact, since $z_{\mathrm{max}}(\Omega_0=1.04)=2575$ and $z_{\mathrm{max}}(\Omega_0=1.01)=6.6 \times 10^6$, we see that  this constraint is never a significant limitation on the bounds.

\end{itemize} 
Note that as $\Omega_0 \to 1^+$, $z_\mathrm{max} \to +\infty$, and equation (\ref{dPSEC}) can be developed in a Taylor series as
\begin{eqnarray}
\fl
d_{P_{\SEC}} (z)&=& \frac{c}{H_0}\;  \ln{\left(1+z\right)} - \frac{c }{6 H_0} \; \left[ \ln{\left(1+z\right)} \right]^3\; \left( \Omega_0-1\right) 
+ O\left([ \Omega_0-1]^2\right).
\end{eqnarray}

If instead one performs a low-redshift expansion, then for general $\Omega_0$
\begin{equation}
d_{P_{\NEC}}(z)  = {c z\over H_0} \left\{ 1 - {z\over2} + \mathcal{O}(z^2) \right\}.
\end{equation}

%----------------
\item[DEC:] Remember that to satisfy the DEC, the Hubble function needs to satisfy both the NEC, inequality (\ref{H_NEC}), \emph{and} the second inequality (\ref{H_DEC}). As a consequence, in order for the DEC to hold, Peebles' angular diameter distance must satisfy inequality (\ref{dP_NEC}), \emph{and} a second inequality to be derived below.
From equation (\ref{H_DEC}), we obtain
\begin{eqnarray}
\fl
J = \int_0^z {{H_0\over H(z)} \; \d z } & \leq & J_{\DEC}\;\equiv\; \int_0^z \frac{\d z}{(1+z)  \sqrt{1+ \Omega_0\; \left[ (1+z)^{4} - 1\right]   }} .
\end{eqnarray}
This integration is a bit more tricky than the previous integrations for the NEC and SEC.\footnote{Indeed, common symbolic manipulation systems such as {\sf Maple} or {\sf Mathematica} require significant human intervention before they will even condescend to verify these results.}
We obtain:
\begin{eqnarray}
\fl
 J_{\DEC}&=& \frac{1}{2\sqrt{1-\Omega_0}} \;
 \ln{ \left\{  \frac{\left( 1-\Omega_0+\sqrt{1-\Omega_0} \right) \left( 1+z\right)^2}{1-\Omega_0+\sqrt{1-\Omega_0}\sqrt{1+\Omega_0 \left[ \left( 1+z\right)^4-1 \right]}  }\right\} } .
\end{eqnarray}

In the general case, we obtain the lower bound on the angular diameter distance,
 \begin{eqnarray}  
 \fl
 d_P(z)\;  \geq \;d_{P_{\DEC}}(z) \equiv \frac{c}{H_0 (1+z)} 
\sqrt{ \frac{\sqrt{1+ \Omega_0 \left[ \left( 1+z\right)^4 -1 \right]} -\left( 1+\Omega_0\left[ \left( 1+z\right)^2-1 \right]\right)}
{2 \Omega_0 \left( 1-\Omega_0 \right)} }; && 
 \nonumber 
 \\
\qquad\qquad \left\{
\begin{array}{ll}
     \forall \; \Omega_0 \; \leq \; 1, \; \forall \; z \; \in [0,+\infty] ;\\
     \forall \;  \Omega_0 \; > \;  1, \; \forall \; z \; \in (z_{\DEC},+\infty] .\\
\end{array} \right.  
&  &
\label{dPDEC}
\end{eqnarray} 
(Equations (19) and (20) of~\cite{santos2} can be viewed as intermediate stages in deriving this result.)
Note that contrast  to the situation for the SEC, there is no constraint on a maximum value for $z$ coming from the requirement that the sine function be monotonic.

The lower bound of the DEC in equation (\ref{dPDEC}) can also be represented in a Taylor series as $\Omega_0 \to 1$,
\begin{eqnarray}
\fl
d_{P_{\DEC}}(z)= \frac{c}{H_0} \frac{ z \left( 2+z \right) }{2 \left( 1+z \right)^2}+ \frac{c}{H_0} \frac{ z^2 \left( 2+z \right)^2\left( 3z^2+6z+4 \right)}{16 \left( 1+z \right)^6}  \left( \Omega_0-1 \right)+O \left( [\Omega_0-1]^2 \right).
\nonumber\\
\end{eqnarray}
If instead one performs a low-redshift expansion, then for general $\Omega_0$
\begin{equation}
d_{P_{\NEC}}(z)  = {c z\over H_0} \left\{ 1 - {(2\Omega_0+1)\; z\over2} + \mathcal{O}(z^2) \right\}.
\end{equation}

\end{description}
%

%-------------------------------------------------------------------------------
\paragraph{Energy conditions and Supernovae data:}
%-------------------------------------------------------------------------------
We can plot the angular diameter distance bounds (NEC, WEC, SEC, and DEC) and compare them with the angular diameter distance raw data from the supernova datasets. We have used data from the supernova legacy survey (\textsf{legacy05})~\cite{legacy, legacy-url} and the Riess \emph{et. al.}  ``gold'' dataset of 2006 (\textsf{gold06})~\cite{gold06}.

%----------------------
\begin{figure}[!htb]
\begin{center}
\includegraphics[width=14cm]{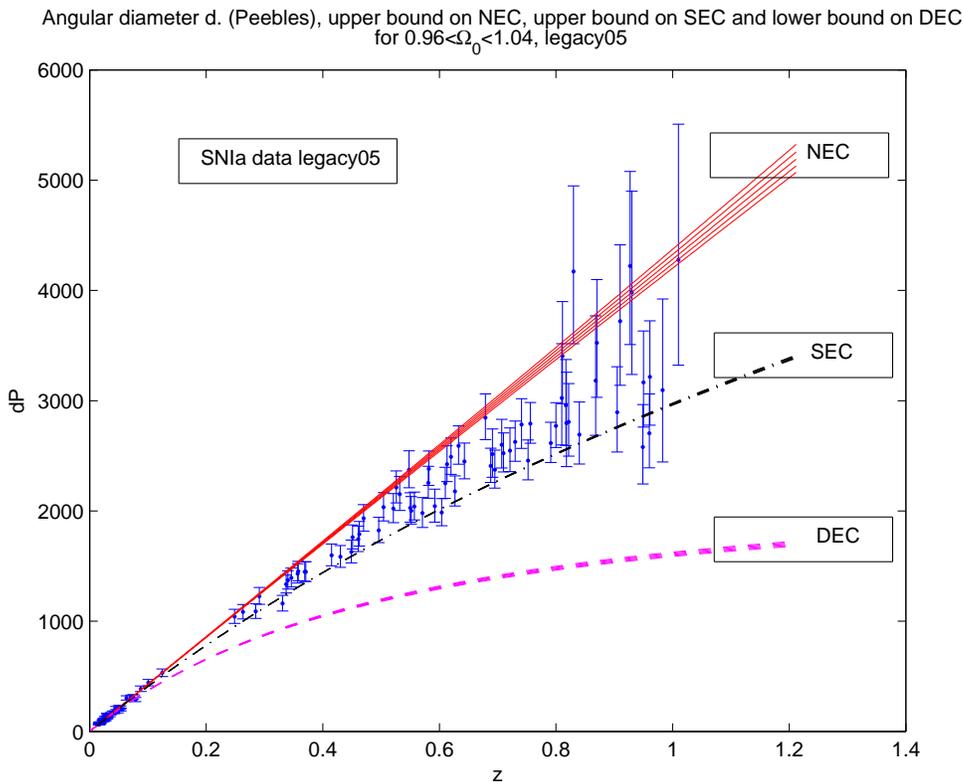}
\end{center}
\caption{\label{F:dP_bounds_legacy05_wider} This figure shows Peebles' angular diameter distance $d_P(z)$ as a function of the $z$-redshift from the nearby and legacy survey, \textsf{legacy05} dataset~\cite{legacy, legacy-url}. Data under the ``red solid lines'' satisfy the NEC/WEC, data under the ``black  dashdot lines'' satisfy the SEC, data under the ``red solid lines'' and above the ``magenta dashed lines'' satisfy the DEC. The 5 lines for each energy conditions correspond to varying values of the parameter $\Omega_0=\{0.96, 0.98, 1.00, 1.02, 1.04 \}$. The value of the Hubble constant is taken to be $H_0=70 \; \mathrm{km/s/Mpc}$.}
\end{figure}
%-----------------------
%----------------------
\begin{figure}[!htb]
\begin{center}
\includegraphics[width=14cm]{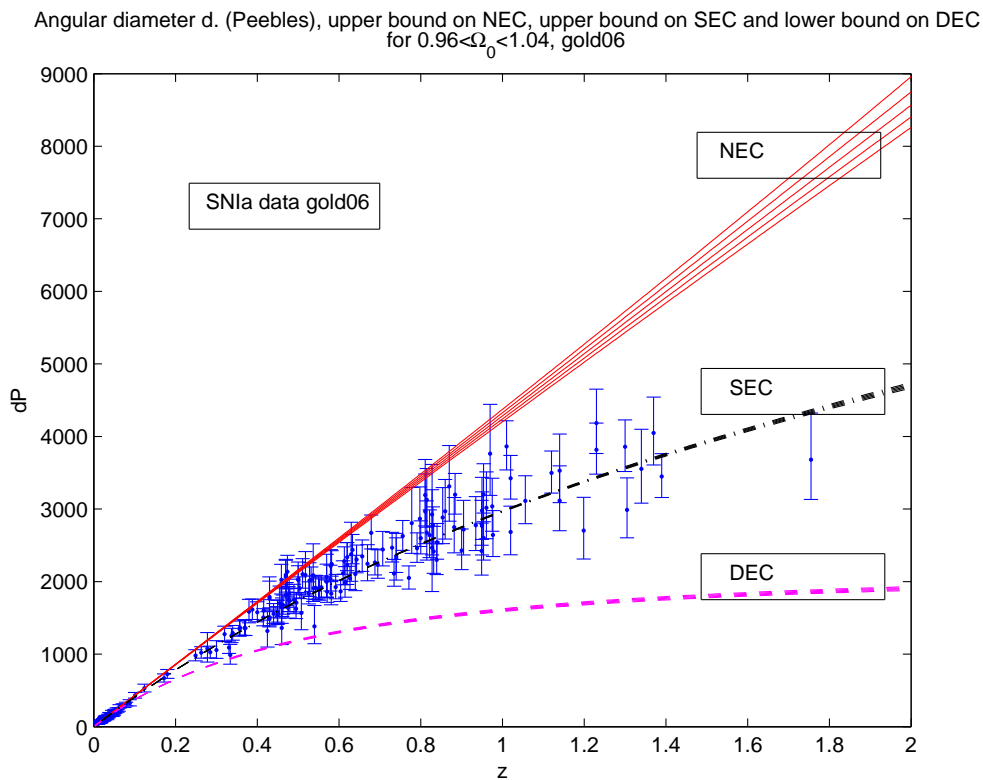}
\end{center}
\caption{\label{F:dP_bounds_gold06_wider} This figure shows Peebles' angular diameter distance $d_P(z)$ as a function of the $z$-redshift from the \textsf{gold06} dataset~\cite{gold06}. Data under the ``red solid lines'' satisfy the NEC/WEC, data under the ``black  dashdot lines'' satisfy the SEC, data under the ``red solid lines'' and above the ``magenta dashed lines'' satisfy the DEC. The 5 lines for each energy conditions correspond to varying values of the parameter $\Omega_0=\{0.96, 0.98, 1.00, 1.02, 1.04\}$. The value of the Hubble constant is taken to be $H_0=70 \; \mathrm{km/s/Mpc}$.}
\end{figure}
%-----------------------

%
\begin{itemize}

\item
Figure \ref{F:dP_bounds_legacy05_wider} compares the upper bounds (NEC and SEC), and the lower bound (DEC), with the \textsf{legacy05} dataset. In contrast figure  \ref{F:dP_bounds_gold06_wider} uses the \textsf{gold06} dataset. 

\item
To satisfy the NEC, the data must be under the ``red solid'' bound, and we can see that most of the data seem to satisfy this condition. 

\item
For the SEC to hold, the data must be under the ``black  dashdot''  bound. Visually it seems ``obvious'' that the data significantly violate the SEC. 

\item
Finally, the DEC is satisfied if both: (1) the NEC is satisfied, and (2)  if the data is above the ``magenta dashed'' lower bound.  This latter condition is well satisfied for the bulk of the data, therefore satisfying the DEC is dependent on the NEC holding.  

\item
As is traditional for estimates of cosmological distance, we plot only one-sigma statistical uncertainties, without any allowance for systematic uncertainties. Any realistic attempt at more careful treatment of the systematics, and/or going to 3-sigma error bars, makes the plots much more disquieting~\cite{Cattoen1,Cattoen2}. 

\item
There seem to be noticeable visual differences when looking at figures  \ref{F:dP_bounds_legacy05_wider}, or   \ref{F:dP_bounds_gold06_wider},  which make it tricky to conclude whether the classical energy conditions are satisfied or not by just looking at the supernova data in isolation. For example, there are a few supernovae data in the redshift range $0.8<z<1$ that appear to violate the NEC in an obvious manner for the \textsf{legacy05} dataset in figure (\ref{F:dP_bounds_legacy05_wider}) . However, the violation does not appear to be as dramatic when looking at the same range of data in the \textsf{gold06} dataset in figure (\ref{F:dP_bounds_gold06_wider}). Another example is that the NEC naively seems to be violated for data in the redshift range $0.4<z<0.6$ in the \textsf{gold06} dataset.   On the contrary, even if  there are less data in the \textsf{legacy05} dataset, one cannot draw the same conclusion.

\item 
Unlike references~\cite{santos1,santos2,santos3}, we believe we cannot draw any conclusions by using the low-redshift linear part of the distance scale curve, as the data has been scaled to enforce a particular value of $H_0$.

\item 
Note in particular that for low redshift the luminosity distance is bounded, \emph{both above and below}, by constraints of the form $c z/H_0 +\mathcal{O}(z^2)$. Thus for data with any statistical uncertainties whatsoever, at low enough $z$, one would \emph{expect} roughly half the supernovae to ``violate'' one or more of these bounds.

\item 
It is important to realise that the ``slope'' of the bounds at $z=0$ depends on the estimate of $H_0$  one adopts (and the precise value of $c$), and can further be affected by the value of the magnitude ``offset'' reported for the data.
\end{itemize}

%------------------------------------------------------------------------------------------
\section{Bounds on the lookback time}
%------------------------------------------------------------------------------------------
The lookback time is defined as~\cite{science,flow,mg8}
\begin{eqnarray}
T(z) &=& \int_a^{a_0} \d t = \int {\d t \over\d a} \;\d a = \int {a\over \dot a} \; {\d a\over a} 
\nonumber \\
 &=& 
 \int {1\over H} \; {\d[a_0/(1+z)]\over a_0/(1+z)} = -\int {1\over H}\;  {\d z/(1+z)^2\over 1/(1+z)} .
\end{eqnarray}
That is
\begin{equation}
T(z) = \int_0^z{1\over (1+z) \; H(z)} \; \d z.
\end{equation}
In order to obtain bounds  on the lookback time for the different energy conditions NEC, WEC, SEC, and DEC, we again use the bounds on the Hubble parameter $H(z)$.
\begin{description}
%----------------------
\item[NEC:] The null energy condition gives a bound on $H(z)$ in equation (\ref{H_NEC}), leading to the inequality
\begin{equation}
\fl
T(z) = \int_0^z{1\over (1+z) \; H(z)} \; \d z  \; \leq \; T_{\NEC}(z)\equiv \int_0^z{1\over (1+z) \; H_{\NEC}(z)}\; \d z 
\end{equation}
We integrate, and substitute the resulting bound into the lookback time (equivalent formulae can be found in reference~\cite{flow}, see also equation (14) of~\cite{santos3})
\begin{equation} 
\fl
T(z) \; \leq \; T_{\NEC}(z)=\frac{1}{H_0 \sqrt{\Omega_0}} \;
\ln{\left[   \frac{\left(1+z \right) \left( 1 +\sqrt{\Omega_0}\right)}
{\sqrt{\Omega_0 +\left( 1-\Omega_0\right)\left( 1+z\right)^2} + \sqrt{\Omega_0}}   \right] } , 
\label{T_NEC}
\end{equation}
this bound is valid $\forall \; \Omega_0 \; \leq \; 1, \; \forall \; z>0$ and $ \forall \;  \Omega_0 \; > \;  1, \; \forall \; z\; \in [0, z_{\NEC}]$.\\
Alternatively, this bound can be rewritten as
\begin{equation} 
\fl
T(z) \leq T_{\NEC}(z)\;=\; \frac{1}{H_0 \sqrt{\Omega_0}} 
\ln{\left[   \frac{\sqrt{\Omega_0+\left( 1-\Omega_0\right)\left( 1+z\right)^2} -\sqrt{\Omega_0} }
{\left(1-\sqrt{\Omega_0}\right)\left( 1+z\right)} \right]} .
\label{T_NEC2}
\end{equation}
Using the standard result that $\sinh^{-1}x = \ln(x+\sqrt{x^2+1})$, we can for $k=-1$ (that is, $\Omega_0<1$) also re-cast this as
\begin{eqnarray}
\fl
T(z) \leq T_{\NEC}(z)\;=\; \frac{1}{H_0 \sqrt{\Omega_0}} 
\left\{ 
\sinh^{-1}\left( \sqrt{\Omega_0\over1-\Omega_0}\right) - \sinh^{-1}\left( {1\over 1+z} \; \sqrt{\Omega_0\over1-\Omega_0}\right)
\right\}.
\nonumber
\\
\label{T_NEC3}
\end{eqnarray}
Equivalent formulae can be found in reference~\cite{flow}. Similarly for $k=+1$ (that is, $\Omega_0>1$) we can use the fact that  $\cosh^{-1}x = \ln(x+\sqrt{x^2-1})$ to obtain
\begin{eqnarray}
\fl
T(z) \leq T_{\NEC}(z)\;=\; \frac{1}{H_0 \sqrt{\Omega_0}} 
\left\{ 
\cosh^{-1}\left( \sqrt{\Omega_0\over\Omega_0-1}\right) - \cosh^{-1}\left( {1\over 1+z} \; \sqrt{\Omega_0\over\Omega_0-1}\right)
\right\}.
\nonumber
\\
\label{T_NEC4}
\end{eqnarray}
(Note that this formula is valid only for $z\leq z_\NEC$, since otherwise the argument of the $\cosh^{-1}$ is less than unity.)
Finally, the upper bound derived from the NEC in equation (\ref{T_NEC}), or equivalently any of equations (\ref{T_NEC2})--(\ref{T_NEC3})--(\ref{T_NEC4}), can also be represented in a Taylor series as $\Omega_0 \to 1$:
\begin{equation}
\fl
T_{\NEC}(z)\;=\; \frac{\ln{\left(1+z\right)}}{H_0}+\frac{z^2+2z-2\ln{\left(1+z \right)}}{4H_0}\left(\Omega_0-1 \right)+O \left([ \Omega_0-1]^2 \right). 
\end{equation}
In particular for $\Omega_0=1$ we recover the results of~\cite{flow} and~\cite{santos3}.

%----------------------
\item[WEC:] This energy conditions gives a bound on $H(z)$ in (\ref{H_WEC}) for $\Omega_0\in(0,1)$ only. Thereby it can be deduced that
\begin{equation}
\fl
T(z) = \int_0^z{1\over (1+z) \; H(z)} \; \d z  \; \leq \; T_{\WEC}(z) \equiv \int_0^z{1\over (1+z) \; H_{\WEC}(z)}\; \d z,
\end{equation}
providing the (weak) bound
\begin{eqnarray}  
T(z) \;\leq\; T_{\WEC}(z)&=& \frac{z}{H_0\;\sqrt{1-\Omega_0} \;\left(1+z\right)}.
\label{T_WEC}
\end{eqnarray} 
Again, this provided no additional useful information beyond the NEC-derived bound.
%----------------------

%----------------------
\item[SEC:] This energy condition gives a bound on $H(z)$  in (\ref{H_SEC}). Thereby it can be deduced (as in the articles~\cite{science, flow, mg8, santos3}) that,
\begin{equation}
\fl
T(z) = \int_0^z{1\over (1+z) \; H(z)} \; \d z  \; \leq \; T_{\SEC}(z) \equiv \int_0^z{1\over (1+z) \; H_{\SEC}(z)}\; \d z,
\end{equation}
that is
\begin{eqnarray}  
T_{\SEC}(z)&=& \frac{z}{H_0 \; \left(1+z\right)}.
\label{T_SEC}
\end{eqnarray} 
The  above result  equation (\ref{T_SEC}) is completely independent of the value of the parameter $\Omega_0$. Note that this result was first introduced by Visser in~\cite{science, flow, mg8}. 
%----------------------

\item[DEC:] Remember that to satisfy the DEC, the Hubble function needs to satisfy the NEC, inequality (\ref{H_NEC}), \emph{and} inequality (\ref{H_DEC}). As a consequence, in order for the DEC to hold, the lookback time must satisfy  inequality (\ref{T_NEC}), \emph{and} a second inequality that we shall derive below.
From equation (\ref{H_DEC}), we obtain
\begin{equation}
\fl
T(z) = \int_0^z{1\over (1+z) \; H(z)} \; \d z  \; \geq \; T_{\DEC}(z) \equiv \int_0^z{1\over (1+z) \; H_{\DEC}(z)}\; \d z,
\end{equation}
that is
\begin{equation}
 T_{\DEC}(z) = {1\over H_0} \int_0^z{1\over (1+z)^2 \; \sqrt{1+\Omega_0\,((1+z)^4-1)}}\; \d z,
\end{equation}
The integration of this bound is considerably harder than for the other energy conditions. Again, common symbolic manipulation systems such as {\sf Maple} or {\sf Mathematica} require significant human intervention before producing anything useful. Let us first write
\begin{equation}
 T_{\DEC}(z) = {1\over H_0 \; \sqrt{\Omega_0}} 
 \int_0^z{1\over (1+z)^4 \; \sqrt{1-(1-\Omega_0^{-1})\,(1+z)^{-4}}}\; \d z,
\end{equation}
and then, (following the procedure of~\cite{flow}), apply the binomial theorem
\begin{equation}
\fl \left[1-(1-\Omega_0^{-1})\,(1+z)^{-4}\right]^{-1/2} = \sum_{n=0}^\infty {-1/2\choose n} \; (-1)^n \; (1-\Omega_0^{-1})^n\;(1+z)^{-4n}.
\end{equation}
Now the binomial series will converge provided
\begin{equation}
\left|  (1-\Omega_0^{-1})\,(1+z)^{-4}\right| < 1,
\end{equation}
and in view of the region we are integrating over, this means that the series for the lookback time will converge provided
\begin{equation}
\left|  1-\Omega_0^{-1}\right| < 1, \qquad\hbox{that is}\qquad \Omega_0\in(1/2,\infty).
\end{equation}
Subject to this condition we can integrate, and obtain the convergent series
\begin{equation}
\fl
 T_{\DEC}(z) = {1\over H_0 \; \sqrt{\Omega_0}} 
 \sum_{n=0}^\infty {-1/2\choose n} (-1)^n {1\over4n+3} (1-\Omega_0^{-1})^n\,\left[ 1- (1+z)^{-4n-3} \right].
\end{equation}
As a practical matter, for many purposes this series representation may be enough, but we can tidy things up somewhat by first defining
\begin{equation}
S(x) =  \sum_{n=0}^\infty {-1/2\choose n} \; {(-x)^n\over4n+3},
\end{equation}
in which case
\begin{equation}
\fl
 T_{\DEC}(z) = {1\over H_0 \; \sqrt{\Omega_0}}  
 \left\{ S\left(1-\Omega_0^{-1}\right) - (1+z)^{-3} \; S\left( {(1-\Omega_0^{-1})\over(1+z)^{4}} \right) \right\}.
\end{equation}
Finally we can recognize that $S(x)$ is a particular example of  hypergeometric series\footnote{
The classical hypergeometric series is given by 
\begin{equation}
_2F_1 \left( a, b; c; x\right) = \sum_{n=0}^{\infty} \frac{(a)_n (b)_n}{(c)_n} \frac{x}{n!},
\end{equation}
where $(a)_n = a (a+1) (a+2) É (a+n-1)$ is the rising factorial, or Pochhammer symbol. The series is in general a convergent power series for values of $x$ such that $|x| < 1$.} 
and so write
\begin{equation}
S(x) =  \sum_{n=0}^\infty {-1/2\choose n} \; {(-x)^n\over4n+3}  = {1\over3} \; _2F_1\left({1\over2}, {3\over4}; {7\over4}; x\right).
\end{equation}
Therefore
\begin{eqnarray}
\fl
 T_{\DEC}(z) &=& {1\over 3 \, H_0 \, \sqrt{\Omega_0}}  \\
 \fl
 &\times &
 \left\{ _2F_1\left({1\over2}, {3\over4}; {7\over4};1 -\Omega_0^{-1}\right) 
 - (1+z)^{-3} \;\; {}_2F_1\left({1\over2}, {3\over4}; {7\over4}; {(1-\Omega_0^{-1}) \over (1+z)^{4}} \right) \right\}.
 \nonumber
\end{eqnarray}
Again this agrees the results reported in~\cite{flow}, and more recently (for $k=0$, $\Omega_0=1$) in~\cite{santos3}. Of course writing the result in terms of hypergeometric functions does not necessarily give one much additional physical insight --- for physical insight the series $S(x)$ is sufficient, and the realization that one is in fact dealing with a hypergeometric function is likely to be useful only if for some reason one wishes to numerically programme the bound into a computer. (The result can also be cast in terms of elliptic integrals~\cite{flow}, but this does not appear to be particularly illuminating.) 

%----------------------
\end{description}
%----------------------

%------------------------------------------------------------------------------------------
\section{Bounds on the Omega parameter $\Omega(z)$}
%------------------------------------------------------------------------------------------
We have the following identity
\begin{eqnarray}
\Omega -1 &=& \frac{k\,c^2}{a^2\,H^2} =  \frac{k\,c^2}{a_0^2\,H_0^2} \; \frac{a_0^2}{a^2}\; \frac{H_0^2}{H^2} 
=\left( \Omega_0-1 \right) \, \left( 1+z \right)^2 \, \frac{H_0^2}{H^2}.
\end{eqnarray}
That is
\begin{equation}
\Omega(z) = 1+\left( \Omega_0-1 \right) \; \left( 1+z\right)^2 \; \frac{H_0^2}{H(z)^2}.
\end{equation}
Therefore, a bound on $H(z)$ automatically implies a bound on $\Omega(z)$.
\begin{description}
%----------------------
\item[NEC:] The null energy condition gives a bound on $H(z)$, as in equation (\ref{H_NEC}), leading to 
\begin{equation}
\Omega_{\NEC} = \frac{\Omega_0}{\Omega_0+\left( 1-\Omega_0\right)\left( 1+z \right)^2}, \label{O_NEC}
\end{equation}
and the inequalities
\begin{eqnarray}
\hbox{if } \Omega_0 <1,& \forall \; z>0,  &\;  \hbox{ then } \Omega(z) \; \geq \; \Omega_{\NEC}; \\
\hbox{if } \Omega_0 =1,& \forall \; z>0,  &\;  \hbox{ then } \Omega(z) \;=\; \Omega_{\NEC} = 1; \\
\hbox{if } \Omega_0 >1, & \forall \; z>0, &\;  \hbox{ then } \Omega(z) \; \leq \; \Omega_{\NEC} .
\end{eqnarray}

Note that as $\Omega_0 \to 1$, equation (\ref{O_NEC}) can be developed in a Taylor series as
\begin{eqnarray}
\Omega_{\NEC} &= &1 +\left( 1+z \right)^2\left(\Omega_0-1 \right)+  O \left( [\Omega_0-1]^2 \right).
\end{eqnarray}
%----------------------

%----------------------
\item[WEC:] The weak energy condition gives a bound on $H(z)$, as in equation (\ref{H_WEC}), but for $\Omega_0\in(0,1)$ only, leading to 
the trivial result  $\Omega_{\WEC} = 0$, and the trivial inequality
\begin{eqnarray}
\hbox{if } \Omega_0 <1,& \forall \; z>0,  &\;  \Omega \; \geq \; \Omega_{\WEC}=0.
\end{eqnarray}
This bound is not useful, except as a consistency check.
%----------------------

\item[SEC:] The strong energy condition gives a bound on $H(z)$, as in equation (\ref{H_SEC}), leading to 
\begin{equation}
\Omega_{\SEC} \equiv\Omega_0,
\end{equation}
and the inequalities
\begin{eqnarray}
\hbox{if } \Omega_0 <1,& \forall \; z>0,  &\;   \hbox{ then } \Omega(z) \; \geq \; \Omega_{\SEC} = \Omega_0; \\
\hbox{if } \Omega_0 =1,& \forall \; z>0,  &\;   \hbox{ then } \Omega(z) \;=\; \Omega_{\SEC}= \Omega_0=1; \\
\hbox{if } \Omega_0 >1, & \forall \; z>0, & \;   \hbox{ then }\Omega(z) \; \leq \; \Omega_{\SEC}=\Omega_0.
\end{eqnarray}
%----------------------
\item[DEC:] The dominant energy condition gives a bound on $H(z)$, as in equation (\ref{H_DEC}), leading to 
\begin{equation}
\Omega_{\DEC} = \frac{\Omega_0 \left( 1+z \right)^4}{1+\Omega_0\left[\left( 1+z \right)^4-1\right]}, 
\label{O_DEC}
\end{equation}
and the inequalities
\begin{eqnarray}
\hbox{if } \Omega_0 <1,& \forall \; z>0,  &\;    \hbox{ then }\Omega(z) \; \leq \; \Omega_{\DEC}; \\
\hbox{if } \Omega_0 =1,& \forall \; z>0,  &\;    \hbox{ then }\Omega(z) \;=\; \Omega_{\DEC} = 1; \\
\hbox{if } \Omega_0 >1, & \forall \; z\; \in [0, z_{\DEC}], &\;    \hbox{ then }\Omega(z) \; \geq \; \Omega_{\DEC} .
\end{eqnarray}

Note that as $\Omega_0 \to 1$, equation (\ref{O_DEC}) can be developed in a Taylor series as
\begin{eqnarray}
\Omega_{\DEC} &= &1 +\frac{(\Omega_0-1) }{\left( 1+z \right)^4} +  O \left( [\Omega_0-1]^2 \right).
\end{eqnarray}
%----------------------

\end{description}
These bounds on $\Omega(z)$ are potentially of interest with regard to cosmological nucleosynthesis, which is effectively sensitive to $\Omega(z_\mathrm{nucleosynthesis})$. More generally, any bound on the number of relativistic particle species at any particular epoch can be converted, with a little work and some technical assumptions, into a bound on the Omega parameter at that epoch.

%------------------------------------------------------------------------------------------
\section{Bounds on the density $\rho(z)$}
%------------------------------------------------------------------------------------------
We have the following identity
\begin{eqnarray}
\rho &=& 3\; \Omega \; H^2 = 3\;\left[ 1+\left( \Omega_0-1\right)\left( 1+z\right)^2 \frac{H_0^2}{H^2} \right] \; H^2.
\end{eqnarray}
That is
\begin{equation}
\rho(z) = 3H(z)^2 +3\left( \Omega_0-1\right)\;\left( 1+z\right)^2 \; H_0^2,
\end{equation}
showing that a bound on $H(z)$ automatically implies a bound on $\rho(z)$.

Alternatively, we can also write the following identity
\begin{eqnarray}
\fl
\rho &=&  3 \left( H^2 + \frac{kc^2}{a^2} \right) = 3H_0^2\; \frac{H^2}{H_0^2}+ \frac{3kc^2}{a_0^2} \; \frac{a_0^2}{a^2} 
= \frac{\rho_0}{\Omega_0} \; \frac{H^2}{H_0^2}+\rho_0 \left( 1-\frac{1}{\Omega_0} \right) \left( 1+z\right)^2.
\end{eqnarray}
That is
\begin{equation}
\rho(z) = \rho_0 \left[   \frac{1}{\Omega_0} \; \frac{H(z)^2}{H_0^2}+   \left( 1-\frac{1}{\Omega_0} \right)\,\left( 1+z\right)^2 \right].
\end{equation}
Again, a bound on $H(z)$ automatically implies a bound on $\rho(z)$.
\begin{description}

%----------------------
\item[NEC:] The null energy condition gives a bound on $H(z)$, as in equation (\ref{H_NEC}), leading to 
\begin{equation}
\rho_{\NEC} =3\Omega_0 \;H_0^2 = \rho_0, 
\label{rho_NEC}
\end{equation}
and the inequality,
\begin{eqnarray}
\forall \; \Omega_0 ,& \forall \; z>0,  &\;  \rho(z) \; \geq \; \rho_{\NEC} = \rho_0.
\end{eqnarray}
This inequality was also derived by more direct means in~\cite{science, flow}.

%----------------------
\item[WEC:] The weak energy condition gives a bound on $H(z)$, as in equation (\ref{H_WEC}), leading to 
\begin{equation}
\rho_{\WEC} =0, 
\label{rho_WEC}
\end{equation}
and the inequality,
\begin{eqnarray}
\forall \; \Omega_0<1 ,& \forall \; z>0,  &\;  \rho \; \geq \; \rho_{\WEC}=0.
\end{eqnarray}
Of course, since by assuming the WEC we have already assumed that $\rho>0$, this bound is not very useful (and is at best a consistency check on the formalism).

%----------------------
\item[SEC:] The strong energy condition gives a bound on $H(z)$, as in equation (\ref{H_SEC}), leading to 
\begin{equation}
\rho_{\SEC} =3 \Omega_0 \; H_0^2  \left( 1+z \right)^2= \rho_0 \; \left( 1+z \right)^2, 
\label{rho_SEC}
\end{equation}
and the inequality,
\begin{eqnarray}
\forall \; \Omega_0 ,& \forall \; z>0,  &\;  \rho(z) \; \geq \; \rho_{\SEC}= \rho_0 \; (1+z)^2.
\end{eqnarray}
This inequality was also derived by more direct means in~\cite{science, flow}.

%----------------------
\item[DEC:] The dominant energy condition gives a bound on $H(z)$ in equation (\ref{H_DEC}), leading to 
\begin{equation}
\rho_{\DEC} = 3 \Omega_0 \; H_0^2  \; \left( 1+z \right)^6= \rho_0  \;\left( 1+z \right)^6,
\label{rho_DEC}
\end{equation}
and the inequality
\begin{eqnarray}
\forall \; \Omega_0, & \forall \; z>0,  &\;  \rho \; \leq \; \rho_{\DEC} = \rho_0  \;\left( 1+z \right)^6.
\end{eqnarray}
This inequality was also derived by more direct means in~\cite{science, flow}.

%----------------------

\end{description}
Note that bounds on the density and the Hubble function are intimately related. A bound on one will automatically provide a bound on the other, and comments made regarding the Hubble bounds can be carried over to this situation as well.

%------------------------------------------------------------------------------------------
\section{Bounds on the pressure $p(z)$}
%------------------------------------------------------------------------------------------
For the pressure $p(z)$ things are a little different; we have the following identity involving the second time derivative of the scale factor:
\begin{eqnarray}
p &=& - \frac{\dot{a}^2}{a^2}-\frac{k\,c^2}{a^2}-2 \frac{\ddot{a}}{a}.
\end{eqnarray}
But
\begin{eqnarray}
   \frac{\dot{a}^2}{a^2}+2 \frac{\ddot{a}}{a} &=&\frac{1}{\dot{a}a^2} \frac{\d(\dot{a}^2a)}{\d t}=\frac{1}{a^2} \frac{\d(\dot{a}^2a)}{\d a}=\frac{1}{a^2} \frac{\d [H^2a^3]}{\d a},
\end{eqnarray}
implying
\begin{equation}
p = -\frac{1}{a^2}\left(  \frac{\d [H^2a^3]}{\d a}+k\,c^2 \right).
\end{equation}
Here the point is that one would need a bound on the \emph{derivative} of $H(z)$ in order to get a direct bound on the pressure $p(z)$. This does not appear to lead to anything useful.

However, if one has a bound on $H(z)$ and hence $\rho(z)$, one can indirectly get a constraint on $p(z)$ via the classical energy conditions. Again, this does not appear to lead to anything useful.

%---------------------------------------------------
\section{Conclusions}
%---------------------------------------------------

In this article we have extended the discussion of~\cite{science,flow,mg8}, and more recently of~\cite{santos1,santos2,santos3}, to develop a number of rugged  and general energy-condition-induced bounds on various cosmological parameters, bounds which have all taken the form
\begin{equation}
X(z) \gtrless X_\mathrm{bound} \equiv X_0  \; f(\Omega_0,z),
\end{equation}
where $X(z)$ is some cosmological parameter, $X_0$ is its present-day value, and $f(\Omega_0,z)$ is a dimensionless function depending on the particular bound under consideration. While most of the bounds we have considered can be derived by ``elementary'' means, in at least two cases the integrals (and other algebraic manipulations) were sufficiently tricky that common  symbolic manipulation programmes required considerable human intervention in order to obtain useful results.

Overall, we have now managed to place bounds of the above form on the Hubble parameter, the Omega parameter, the density, the lookback time, and various cosmological distances --- where for simplicity of presentation we have focussed on Peebles' angular diameter distance as being representative. We have briefly sketched how to use these bounds to confront the supernova data, but have not yet performed any detailed analysis of this point. 

%---------------------------------------------------
\appendix
%---------------------------------------------------

%------------------------------------------------
\ack
%------------------------------------------------

This Research was supported by the Marsden Fund administered by the Royal Society of New Zealand. CC was also supported by a Victoria University of Wellington Postgraduate scholarship.

%------------------------------------------------
\section*{References}
%------------------------------------------------

%------------------------------------------------

%--------------------------------------------------- 
\end{document}